\documentclass[11pt,letterpaper,fleqn]{article}
\usepackage[hyphens]{url}
\usepackage{hyperref}
\newcommand{\reportnumber}{FERMILAB-PUB-26-0747-CSAID-T}
\usepackage[angle=0,scale=1,color=black,firstpage=true,opacity=1]{background}
\SetBgContents{\footnotesize\sf{\reportnumber}}
\SetBgPosition{current page.north east}
\SetBgHshift{-2in}
\SetBgVshift{-0.5in}

\usepackage[backend=biber,style=numeric-comp,sorting=none,eprint=true]{biblatex}
\appto\biburlsetup{%
}

\AtBeginBibliography{\raggedright}

\usepackage{xcolor}
\usepackage{enumitem}
\usepackage[sf,bf,compact,topmarks,calcwidth,pagestyles]{titlesec}
\usepackage{authblk}
\usepackage{parskip}
\usepackage[margin=1in,headheight=0pt,headsep=0pt]{geometry}

\selectcolormodel{rgb}

\setlist[itemize]{itemsep=1pt, parsep=1pt, before=\setlength{\parskip}{1pt}}

\titleformat{\section}[block]{\normalfont\scshape}{\thesection}{0.5em}{\scshape}
\titleformat{\subsection}[block]{\normalfont\scshape}{\thesubsection}{0.5em}{\scshape}
\titleformat{\subsubsection}{\normalfont\scshape}{\thesubsubsection}{0.5em}{}
\titleformat{\paragraph}[runin]{\normalfont\scshape}{\theparagraph}{0.5em}{}
\titleformat{\subparagraph}[runin]{\normalfont\scshape}{\thesubparagraph}{0.5em}{}

\title{{\bf Blueprint for an AI-Enabled Particle Theory Community}}
\date{\today}

\DeclareRobustCommand{\secondaryauthors}{%
  \par\vspace{1.5ex}%
  {\color{gray}\rule{0.3\textwidth}{0.4pt}}\par\vspace{1ex}%
  \footnotesize
  \mbox{Prateek Agrawal$^{9}$},
  \mbox{Jonathan Bagger$^{10}$},
  \mbox{Yang Bai$^{11}$},
  \mbox{Brian Batell$^{12}$},
  \mbox{Sean Benevedes$^{13}$},
  \mbox{Vedran Brdar$^{14}$},
  \mbox{Matthew R.\ Buckley$^{6}$},
  \mbox{Christopher Carone$^{15}$},
  \mbox{Mu-Chun Chen$^{16}$},
  \mbox{Kun Cheng$^{12}$},
  \mbox{So Chigusa$^{7}$},
  \mbox{R.~Sekhar Chivukula$^{17}$},
  \mbox{Neil Christensen$^{18}$},
  \mbox{Timothy Cohen$^{19}$},
  \mbox{Jonathan M.\ Cornell$^{20}$},
  \mbox{Antonio Delgado$^{21}$},
  \mbox{Radovan Dermisek$^{22}$},
  \mbox{Jeff A.\ Dror$^{23}$},
  \mbox{Anatoly Dymarsky$^{24}$},
  \mbox{Jonathan L.~Feng$^{16}$},
  \mbox{Terrance Figy$^{25}$},
  \mbox{Alexander Friedland$^{5}$},
  \mbox{Dorival Goncalves$^{14}$},
  \mbox{Craig Group$^{26}$},
  \mbox{Tao Han$^{12}$},
  \mbox{Mark Hertzberg$^{27}$},
  \mbox{Samuel Homiller$^{12}$},
  \mbox{Matheus Hostert$^{28}$},
  \mbox{Philip Ilten$^{8}$},
  \mbox{Zhong-Bo Kang$^{29}$},
  \mbox{Can Kilic$^{30}$},
  \mbox{Doojin Kim$^{31}$},
  \mbox{William H.\ Kinney$^{32}$},
  \mbox{Simon Knapen$^{33}$},
  \mbox{Kyoungchul Kong$^{34}$},
  \mbox{Graham Kribs$^{19}$},
  \mbox{Shirley Li$^{16}$},
  \mbox{Huey-Wen Lin$^{35}$},
  \mbox{Zhen Liu$^{36}$},
  \mbox{Ian Low$^{37}$},
  \mbox{Pedro Machado$^{1}$},
  \mbox{Kendall Mahn$^{35}$},
  \mbox{Alexander Maloney$^{38}$},
  \mbox{Liam McAllister$^{39}$},
  \mbox{Djordje Minic$^{40}$},
  \mbox{Bernhard Mistlberger$^{5}$},
  \mbox{Sebastian Mizera$^{41}$},
  \mbox{Moritz M\"unchmeyer$^{11}$},
  \mbox{Ethan Neil$^{42}$},
  \mbox{Nobuchika Okada$^{4}$},
  \mbox{Sofia Palacios Schweitzer$^{6}$},
  \mbox{Ian Pang$^{1,4}$},
  \mbox{Andreas Papaefstathiou$^{43}$},
  \mbox{Alexey Petrov$^{44}$},
  \mbox{David Poland$^{45}$},
  \mbox{Xiaolin Qi$^{40}$},
  \mbox{Michael Ratz$^{16}$},
  \mbox{Laura Reina$^{46}$},
  \mbox{Thomas Rizzo$^{5}$},
  \mbox{Nicholas Rodd$^{33}$},
  \mbox{Alessandro Russo$^{47}$},
  \mbox{Philip Schuster$^{5}$},
  \mbox{Phiala Shanahan$^{7}$},
  \mbox{Elizabeth H.\ Simmons$^{17}$},
  \mbox{George Siopsis$^{48}$},
  \mbox{Dejan Stojkovic$^{32}$},
  \mbox{Shufang Su$^{49}$},
  \mbox{Tim M.\,P.\ Tait$^{16}$},
  \mbox{Lian-Tao Wang$^{50}$},
  \mbox{Zhengkang Zhang$^{51}$}
  \par\vspace{1ex}\scriptsize
  \mbox{$^{9}$\,University of California, Santa Barbara},
  \mbox{$^{10}$\,Johns Hopkins University},
  \mbox{$^{11}$\,University of Wisconsin--Madison},
  \mbox{$^{12}$\,University of Pittsburgh},
  \mbox{$^{13}$\,Georgia Institute of Technology},
  \mbox{$^{14}$\,Oklahoma State University},
  \mbox{$^{15}$\,William \& Mary},
  \mbox{$^{16}$\,University of California, Irvine},
  \mbox{$^{17}$\,University of California, San Diego},
  \mbox{$^{18}$\,Illinois State University},
  \mbox{$^{19}$\,University of Oregon},
  \mbox{$^{20}$\,Weber State University},
  \mbox{$^{21}$\,University of Notre Dame},
  \mbox{$^{22}$\,Indiana University},
  \mbox{$^{23}$\,University of Florida},
  \mbox{$^{24}$\,University of Kentucky},
  \mbox{$^{25}$\,Wichita State University},
  \mbox{$^{26}$\,University of Virginia},
  \mbox{$^{27}$\,Tufts University},
  \mbox{$^{28}$\,University of Iowa},
  \mbox{$^{29}$\,University of California, Los Angeles},
  \mbox{$^{30}$\,University of Texas at Austin},  
  \mbox{$^{31}$\,University of South Dakota},
  \mbox{$^{32}$\,University at Buffalo, SUNY},
  \mbox{$^{33}$\,Lawrence Berkeley National Laboratory},
  \mbox{$^{34}$\,University of Kansas},
  \mbox{$^{35}$\,Michigan State University},
  \mbox{$^{36}$\,University of Minnesota},
  \mbox{$^{37}$\,Northwestern University},
  \mbox{$^{38}$\,Syracuse University},
  \mbox{$^{39}$\,Cornell University},
  \mbox{$^{40}$\,Virginia Tech},
  \mbox{$^{41}$\,Columbia University},
  \mbox{$^{42}$\,University of Colorado Boulder},
  \mbox{$^{43}$\,Kennesaw State University},
  \mbox{$^{44}$\,University of South Carolina},
  \mbox{$^{45}$\,Yale University},
  \mbox{$^{46}$\,Florida State University},
  \mbox{$^{47}$\,Stanford University},
  \mbox{$^{48}$\,University of Tennessee, Knoxville},
  \mbox{$^{49}$\,University of Arizona},
  \mbox{$^{50}$\,University of Chicago},
  \mbox{$^{51}$\,University of Utah}
  \par}

\author{%
\mbox{George T.\ Fleming$^1$}, %
\mbox{Daniel C.\ Hackett$^1$}, %
\mbox{James Halverson$^2$}, %
\mbox{T.~J.~Hobbs$^3$}, %
\mbox{Stefan Hoeche$^1$}, %
\mbox{Brandon Kriesten$^3$}, %
\mbox{Konstantin~T.~Matchev$^4$}, %
\mbox{Tony Menzo$^{1,4}$}, %
\mbox{Stephen Mrenna$^1$}, %
\mbox{Benjamin~Nachman$^{5}$}, %
\mbox{Prasanth Shyamsundar$^1$}, %
\mbox{David Shih$^6$}, %
\mbox{Jesse~Thaler$^7$}, %
\mbox{Jure Zupan$^8$}\\%
$^1$\,Fermi National Accelerator Laboratory, %
$^2$\,Northeastern University, %
$^3$\,Argonne National Laboratory, %
$^4$\,University~of~Alabama, %
$^5$\,SLAC National Laboratory, %
$^6$\,Rutgers University, %
$^7$\,Massachusetts~Institute~of~Technology, %
$^8$\,University~of~Cincinnati%
\secondaryauthors%
}

\begin{document}
\maketitle
\begin{center}
  \begin{minipage}{\textwidth}
  \section*{\centering Abstract}
  Rapid advances in the capabilities and reach of artificial intelligence (AI) are changing
  research in particle physics.
  In this new era, AI is an innovative tool for theoretical physicists, while
  the interaction of AI and particle theory is itself a developing field.
  We argue that particle theory requires a community-scale
  effort to realize its full potential.
  We show how particle theory occupies a unique position in the AI for physics landscape, and
  describe the complementarity of AI and theory developments.
  We conclude with a series of recommendations on how to build an AI-enabled theory community for the Genesis Mission.
  \end{minipage}
\end{center}
\vspace*{1cm}
\setcounter{page}{1}
\section{Response to the National Challenges}
\label{sec:grand-challenges}

This white paper is a response to the Genesis Mission (GM)\footnote{See \url{https://www.energy.gov/undersecretaryforscience/genesis-mission/genesis-mission}.}
from a group of
particle theorists at universities and national laboratories.
The GM aims to stimulate advances in Artificial Intelligence (AI), which includes Machine Learning (ML) and Large
Language Models (LLMs).
Two of the Science and Technology challenges identified by the Genesis Mission (GM)~\cite{GM_ST_challenges},
Unifying Physics from Quarks to Cosmos and
Discovering Quantum Algorithms with AI,
are aligned with the particle theory community:
\begin{itemize}
\item
to integrate the vast array of microscopic to cosmological data with theoretical knowledge to accelerate discovery.
\item
to accelerate the design and development of quantum circuits and algorithms that represent realistic quantum field theories (QFTs) and have provable
 quantum advantages.
\end{itemize}

Our purpose is to highlight the role particle theory plays in the GM and provide a vision how
theorists can help realize the GM goals.   This contribution is the theory counterpart to the
experiment-oriented white paper~\cite{Aarrestad:2026xrs}, which sets out an AI-native research
ecosystem for experimental particle physics.  We take theory and experiment to be two wings of a single national
HEP AI effort rather than parallel enterprises.  However, we argue
that the theory wing is distinct and valuable and requires its own community-wide representation.

\subsection{How Do and Will Theoretical Physicists Use AI}

The GM aims to utilize the latest and quickly developing advances in AI
to accelerate the pace of scientific discovery. Developments within just the last year include: large multimodal models (LMMs) that are
pretrained to receive information in various modalities, e.g.\ text, images, video, audio;
LLMs trained to serve as AI agents via tool calling; and, within just the last \textit{few months},
 ``coding agents'' that carry out nontrivial computational (science) tasks with minimal expert human guidance.
Theoretical physicists are positioned advantageously to use these latest developments
in two broad ways:

\noindent\textit{Theory Building}, which means building physical models based on some theoretical principles, usually encoded in the language of mathematics, and then performing calculations to test if they reproduce known physical results. 
Agentic AI can assist at every stage in this exploratory process at scales and speeds not previously possible~\cite{Bakshi:2025fgx,Menzo:2025cim,Miao:2025sms,Agrawal:2026lvg,Menzo:2026qrl,Faroughy:2026dkj,Saad:2026pan}.

\noindent\textit{Theory Discovery}, 
a form of inverse problem,
is analyzing large physical datasets 
to infer what underlying theories or models could have produced them
and potentially discover new physical principles. 
Agentic AI has demonstrated its utility in exploring large datasets~\cite{Gendreau-Distler:2025fsj,Moreno:2026mqk,Badea:2026klb,He:2026jjb,Diefenbacher:2025zzn}.
Furthermore, bespoke physics-specific LMMs can be pre-trained on large 
experimental or observational physics datasets and then mined for features that point to the most important
aspects of the data that any good theoretical model must reproduce. 

Particle theory serves an important and complementary role to experimental
particle physics.%
\footnote{Broadly speaking, theoretical physics~\cite{Craig:2022cef}:
motivates experiments and experimental facilities by proposing explanations
for data that can be confirmed or falsified;
interprets experimental results by connecting observations to abstract mathematical structures;
suggests new experiments or analyses based on the predictions of models or theories;
explores the consequences of theories and models in future experiments;
extends the boundaries of our understanding of Nature and incorporates new perspectives
(such as quantum information);
develops new techniques relevant to experiments and builds tools used by the experimental community for simulation and analysis;
and constructs theories and models that connect particle physics to gravity, cosmology, astrophysics, nuclear physics, condensed matter, atomic, molecular, and optical (AMO) physics, and mathematics.}
Theory building and discovery are its bread-and-butter.
More specifically,
particle theory has many specialized requirements that are not well suited to ``off-the-shelf'' AI applications, thereby
driving the discovery of new ways to use AI.
Theoretical physicists have contributed to many of the recent advances in fundamental AI, acquiring non-traditional skills
in data science, modern applied statistics, AI/ML concepts and architectures.
Moreover, particle theory can
generate its own synthetic data that can be improved systematically and regenerated
under different theoretical assumptions.
Theory can both supply the data from which
patterns are inferred and subject those inferences to systematic verification and
validation, establishing representations
and standards before they are needed at the HL-LHC, DUNE, the EIC, and beyond. 

\subsection{The Need for a Theory AI Community}
\label{sec:need-for-community}

Experimental collaborations have O(1000) members for a very good reason:  a detector for a frontier experiment
cannot be built and operated by a small group, and the detector provides valuable data, a joint asset shared for
semi-independent research.  To achieve their ends, the collaborations adopt structure and hierarchy that
provide direction and oversight, enforce standards, evaluate the rigor and validity of scientific results and manage resources.

This scale of organization does not exist in the theory community.
Even a well-funded research team at an academic institution working in theory might consist of at most a faculty advisor, a postdoctoral researcher and several graduate students (experimental groups can be only slightly larger, but they self-organize into larger research groups within collaborations).  For the most part (see the USQCD example below), any larger scale organization of particle
theory has
not been viewed as necessary, and the independence of theoretical physics has been viewed as positive.  However, the potential of
AI-driven theory requires us to reconsider what is truly fundamental to theoretical physics.

Organizing at community scale would unlock assets that no individual researcher or research group can obtain:
\begin{itemize}
\item \textbf{Frontier-scale computing resources and model access:} resources beyond a single PI's budget and not subject to the non-uniformity of university resources.
\item \textbf{The agent-interaction corpus:}
a collective record of how agents attack theory problems, including prompts, tool calls, generated code, failures, and human corrections which, when pooled across many groups, can be used to improve future agentic harnesses with proper human oversight to improve scientific alignment.
Theorists have a direct scientific interest in the problem-solving strategies chosen by LLMs.
\item \textbf{Shared validation infrastructure:} benchmarks for agentic theory workflows needed to build trust, and to establish which models suffice for which tasks. This is only useful if a wide variety of use cases are available \cite{Chung:2025nsd,Zhu:2025qnm,Hawks:2025lkm,Faroughy:2026dkj,Miao:2026gmx}.
\item \textbf{Scientific foundation models:} training runs no single group can fund, whose value scales with the diversity of contributed data.
\end{itemize}

These assets did not exist five years ago, and there was no reason to, or mechanism for, the theory community to organize.
With the demonstrated power of state-of-the-art AI and the launch of the GM, this situation has changed.
To obtain these goods, the theory community needs to bind into a larger, organized and coherent structure.

Also, the current paradigm of
``reinforcement learning with verifiable rewards'' has led to rapid advances in agentic capabilities but also well-publicized incidents of agents exhibiting misaligned behaviors including cheating \cite{Baker:2025monitoring,wang2026rewardhackingeralarge}, unintended collusion with other agents \cite{motwani2025secretcollusionaiagents,fish2026algorithmiccollusionlargelanguage}, breaking out of sandboxes \cite{meinke2025frontiermodelscapableincontext,greenblatt2024alignmentfakinglargelanguage}, illegal hacking \cite{Anthropic:2025espionage}, \textit{etc.} One explanation is insufficient human oversight of powerful agents. In the theory domain, engaging with the broader theory community leverages human resources necessary to make sure theory agents remain properly aligned with scientific practices.

An analogous situation occurred over three decades ago in the rise of high performance computing (HPC) and the impact on a major consumer in theoretical HENP, lattice gauge theory. Technology was evolving rapidly with software co-evolving at the same pace and resources becoming obsolete every few years, beyond the capabilities of individual universities and small research groups to keep up.  The U.S. lattice field theory community banded together to form the USQCD organization to coordinate efforts on keeping up with the latest advances in technology (hardware and software) while leaving researchers free to pursue their individual theoretical physics interests using the shared community technology.  The mere fact that the organization has persisted and thrived for over two decades when other national theory-oriented collaborations have come and gone is evidence enough that it is a collaborative model for theorists worth emulating.

The organization of an HEP-experimental AI community (as outlined in \cite{Aarrestad:2026xrs})
is already underway.  Theory should not duplicate that effort; we argue that theory AI research
differs enough from its experimental counterpart to warrant its own organizational unit,
even where the underlying infrastructure is best shared.
A critical part of theory research in AI will be keeping the traditional, smaller groups connected to a larger effort \textbf{and} sharing what they learn, reporting both research successes as well as putative failures.

\section{Four Modes of AI in Particle Theory}
\label{sec:ai-in-theory}

The following subsections detail the way theoretical physics can benefit from and contribute to the GM:
agentic AI to accelerate standard
theory tasks (Section~\ref{sec:agentic-ai});
theory of and for AI to better understand why and how AI
works and develop better applications (Section~\ref{sec:theory-of-ai});
bespoke AI for the unique problems that
particle theory aims to solve (Section~\ref{sec:bespoke});
and the construction and study of
scientific foundation models that could lead to a deeper understanding of
the structure of nature (Section~\ref{sec:foundation-models}).
The shared assets identified in Section~\ref{sec:need-for-community}
are critical to these developments:
agentic AI is the principal consumer of frontier model access and
the source of the agent-interaction corpus;
the shared validation infrastructure can establish which models are
trustworthy for which theory tasks.
Frontier-scale computing resources, meanwhile, are needed to train scientific foundation models.
Theorists not only benefit from AI but also contribute uniquely to its
further development.
The subject of what theory contributes is addressed in Section~\ref{sec:theory-of-ai}.

\subsection{Agentic AI in Theoretical Physics}
\label{sec:agentic-ai}

AI agents can take on different roles, from 
\textbf{supervised research assistant}, to
\textbf{autonomous scientific assistant}, to
\textbf{autonomous scientist(s)},
\textit{etc.}
Here, we outline how these different agents can accelerate particle theory and AI advances.

\paragraph{AI Supervised Research Assistant}
The AI supervised research assistant does nontrivial work in
close interaction with its human operators, who provide guidance, feedback, and
verification.

There are various early examples of theoretical physics results produced in this manner,%
\footnote{Further examples in pure mathematics include solutions to some of the Erd\H{o}s problems~\cite{Feng:2026erdos}, the First Proof Challenge problems~\cite{Abouzaid:2026firstproof}, and results in graph combinatorics~\cite{Wagner:2021comb,Novikov:2025alphaevolve}.}
such as the AI-guided discovery of an unexpected connection between different invariants of knots, producing a proven
theorem~\cite{knot_nature}, closed-form expressions for certain gluon amplitudes~\cite{Guevara:2026qzd}, a resummation of large logarithms in collider event shapes~\cite{Schwartz:2026ekw}, and new ML methods for symbolic simplification of scattering amplitudes and Feynman loop integrals~\cite{Cheung:2024svk,Cai:2024znx,Shih:2026lmy,Shih:2026jfe,
Tait:2026guk,Tait:2026vrd}.

Emerging tools are further expanding what is possible for theory-oriented AI. AI proof
assistants can verify formal mathematics, helping
theorists construct and check rigorous proofs interactively~\cite{mathlib,Hubert:2026alphaproof}.  LLMs
can assist with direct symbolic calculation, reshaping
conventional computational pipelines in formal theory~\cite{Alnuqaydan:2022ncd,Menzo:2026qrl,Saad:2026pan,Gu:2026zbf}.
An ``AI co-scientist'' is becoming a practical reality, capable of generating hypotheses from
theoretical data, which the human scientist evaluates, selects, and
develops~\cite{Gottweis:2025coscientist,Lu:2024aiscientist,Miao:2025sms,Bakshi:2025fgx}.

\paragraph{AI Autonomous Scientific Assistant}

Much of our day-to-day work involves applying
well-established methodology to carry out complex but predictable tasks
requiring minimal novel insight to complete.  This includes routine mathematical
calculations, managing and running
simulations,
and basic data
analysis~\cite{Menzo:2025cim,Qiu:2026iby,Plehn:2026gxv,Faroughy:2026dkj,Gendreau-Distler:2025fsj,Desai:2026nmx,Jiao:2026dsu,Gao:2026cpd,Wang:2026cof}.

Many of these tasks can be automated with current or near-term AI, including:
  (1) calculations across families of new physics models and associated experimental searches to set limits and optimize
    searches~\cite{Saad:2026pan,Xiao:2026gal,Guo:2026ifi,Diefenbacher:2026azr,Lucente:2026kgh,Wang:2026jjn};
  (2) tests of different string theories and the exploration of the large
    parameter space of the string landscape~\cite{Ruehle:2020jrk,Halverson:2019tkf,Matchev:2024ash};
  (3) searches over quantum circuit designs, quantum algorithms, and related
    constructions~\cite{Ruiz:2024boz,Cao:2025shc,Knipfer:2026kng};
  (4) optimization of algorithmic hyperparameters; and
  (5) orchestration of computing campaigns~\cite{Menzo:2025cim,Roman:2026rcn,Lugato:2026osi}.

\paragraph{AI Autonomous Scientist(s)}

This use case envisions 
AI acting as autonomous scientists, or swarms of them, attacking nontrivial research problems~\cite{Lu:2024aiscientist,Miao:2025sms,Tan:2026ier,Alexander:2026lpw}.
At its most powerful extreme, this may include closed-loop systems generating and exploring novel hypotheses. It is critical to investigate their viability, but realizing automated science workflows will require significant
experimentation.
Within
this pursuit, there are two interrelated approaches:
long-running autonomous ``stateful'' agents that maintain context and
  build on their own prior work over extended periods~\cite{Kwa:2025metr,Miao:2025sms}; and
multi-agent orchestration, where multiple AI agents are deployed with
  complementary roles~\cite{Roman:2026rcn,Plehn:2026gxv,He:2026jjb,Lucente:2026kgh,Birk:2026zpd,Costa:2026oew,Hammad:2026ged,Saito:2026tfq}.
Both sit at the frontier of modern AI engineering, undergoing
extensive exploration across science and technology.
A natural testing ground for such
systems is \textbf{formal theory},
where theoretical data arises from definitions and can be both generated and validated without
physical experiments~\cite{Gu:2026zbf,Menzo:2026qrl}.

\subsection{Theory of and for AI}
\label{sec:theory-of-ai}

The relationship between theoretical physics and AI is bi-directional.
AI can accelerate the study of physics, while the mathematical tools of physics can
deepen our understanding of AI itself.
Particle theorists
have a long history of exploring and explaining complex mathematical
structures.  QFTs provide both
good problems and good datasets for understanding how and why AI/ML works.

\paragraph{Understanding AI/ML through Physics}

At its core, ML is the science of constructing, optimizing, and
using function approximations, most often via stochastic optimization.  The
mathematics underlying ML is statistics on function spaces, closely related
to QFT and statistical mechanics, and the same mathematical
tools developed for QFT can be applied to study AI/ML~\cite{Roberts:2021fes,Halverson:2024hax}.
Physical reasoning can also shed light on learning dynamics and improve
interpretability.

Neural network quantum field theory (NNQFT)~\cite{Halverson:2020trp,Demirtas:2023fir} sits at the
intersection of understanding and advancing both AI and physics.  NNQFT 
provides constructive definitions of QFTs, offering new insight and
a fundamentally new approach to QFT itself~\cite{Ferko:2026axm}.

A powerful application of QFT tools to AI/ML is for uncertainty quantification,
an open challenge of critical
importance across the sciences.  
In this context,
QFT tools allow us to characterize the priors and biases encoded
by different ML setups.

\paragraph{Advancing AI/ML with Physics}

Beyond understanding existing methods, there is the potential to advance AI/ML with
new physics-inspired frameworks and architectures.
Diffusion models, the state-of-the-art for image generation
and an emerging approach to language modeling, are defined in terms of
stochastic processes, the same
mathematics behind Brownian motion~\cite{Sohl-Dickstein:2015dhe,Ho:2020epu,Nie:2025llada}.
Innovations in the theory of ML, such as the development of
symmetry-equivariant architectures driven by theoretical work in lattice
field theory and collider phenomenology, give measurable gains in lattice gauge
theory sampling and may have transformative impact on other
scientific computations~\cite{Albergo:2019eim,Kanwar:2020xzo,Boyda:2020hsi,Gong:2022lye,Qu:2022mxj}.

\subsection{Bespoke AI/ML for Physics}
\label{sec:bespoke}

High-quality learned approximations in theoretical physics, derived with custom AI/ML,  can be useful for
learning structure from ``theoretical'' and experimental data, accelerating
simulations and sampling, and constructing theory-informed models for enhanced
sensitivity or inference.
Agentic AI can itself be used to accelerate progress in this area~\cite{Esmail:2026jpb,Diefenbacher:2025zzn,Gu:2026zbf}.

\paragraph{Learning from Theoretical and Physics Data}

AI/ML can help explore new facets of pure mathematics and formal
theoretical physics, including QFT and string theory.  This exploration  
is characterized by manipulations of symbolic formulae or
``theoretical data'': data arising from definitions that is obtained via
mathematical classification or algorithmic generation.  Examples include
fundamental objects of abstract algebra, topology, and differential geometry:
groups and their representations, topological spaces and their invariants, and
metrics with special characteristics.  Such data is of fundamental mathematical
interest, but also of physical interest due to their role in defining
physics theories or their associated observables.  

Early, concrete achievements include the use of neural
networks to uncover hidden mathematical relations in string theory
data~\cite{Carifio:2017bov,He:2017aed,Krefl:2017yox,Ruehle:2017mzq}, driving
conjecture generation via interpretability that led to rigorous new theorems.
Further opportunities exist in computing Calabi-Yau metrics, Kaluza-Klein
modes, and Yukawa couplings via learned methods~\cite{Anderson:2020hux,Douglas:2020hpv,Jejjala:2020wcc}, as well as in applying ML to
QFT amplitudes~\cite{Alnuqaydan:2022ncd,Gu:2026zbf}.
More broadly, symbolic regression, data mining, and related techniques offer
promising avenues for extracting useful patterns from physics data and
discovering new functional relationships~\cite{Udrescu:2019mnk,Dong:2022trn,Cranmer:2023pysr,Matchev:2024ash}.

\paragraph{Accelerating Simulations and Sampling}

Particle physics benefits from sophisticated numerical simulations of particle collisions, lattice field theories, cosmology, and more; in practice, discoveries, measurements, and limits are based on
comparisons of simulator predictions to data.  AI/ML methods can address the significant computational bottlenecks that limit the use and further development of these simulation tools~\cite{Feickert:2021ajf,Shanahan:2022ifi,Butter:2022rso}.
Progress has already been seen in:
  (1) improving speed and efficiency over traditional Monte Carlo methods~\cite{Albergo:2019eim,Bothmann:2020ywa,Gao:2020zvv,Matchev:2020jqz};
    (2) fast and accurate surrogate models for slow physics simulations such as Geant4~\cite{Krause:2024avx};
  (3) integrating parameter differentiability to enable fitting and
   uncertainty estimation~\cite{Heinrich:2022xfa};
  (4) enabling on-the-fly inference~\cite{Cranmer:2019eaq,Kong:2022rnd};
  and (5) providing alternative physics models for less well-known physics~\cite{Ilten:2022jfm,Chan:2023ume,Bierlich:2023zzd,Assi:2025avy,Butter:2025wxn}.

\paragraph{Theory-Informed ML Models}

Task-specific ML models can incorporate physics constraints and
assumptions (symmetries, equivariances, new physics theory scenarios) directly
informed by theory.  These may enhance signal sensitivity in searches
beyond what generic methods achieve.

An important application of AI is solving large-scale
inverse problems.  Particle phenomenology both supplies data and is
reciprocally guided by many inverse problems, including effective field
theories~\cite{Brehmer:2018kdj,Ethier:2021bye,GomezAmbrosio:2022mpm}, UV-complete theory parameterizations~\cite{Harvey:2021oue,Saad:2026pan,Alexander:2026lpw}, parton distribution functions and hadron structure~\cite{NNPDF:2021njg,Kriesten:2023uoi,Kriesten:2025gti}, and
hadronization models~\cite{Ilten:2022jfm,Bierlich:2023zzd,Assi:2025avy}.  AI/ML can be closely integrated into the
parametrization and inference tasks in this space, enabling new approaches to
uncertainty quantification, validation, and verification with consistent
treatments of theoretical and experimental uncertainties carried through the
full phenomenology workflow~\cite{Brehmer:2019xox,Brehmer:2020cvb,Cranmer:2019eaq}, including unfolding detector effects \cite{Andreassen:2019cjw} and separating/explaining uncertainty contributions~\cite{Kriesten:2024ist,Kriesten:2024are}.

\paragraph{AI/ML for Quantum Information Science}

AI can help mitigate the theoretical complexity of QFT
and the physical limitations of current quantum
hardware~\cite{Catterall:2022wjq}.  Driven by the bottleneck of classical
computing, a concerted effort is underway to build the theoretical frameworks
and algorithms necessary to simulate QFT on both digital quantum computers and
analog simulators.  Custom AI/ML methods have the potential to accelerate this
effort across multiple fronts, from quantum circuit optimization and algorithm
construction to learned error mitigation strategies~\cite{Ruiz:2024boz,Cao:2025shc,Knipfer:2026kng}.
The community needs to converge on a set of domain-specific benchmarks that can demonstrate
empirical quantum advantage on QFT problems.

\subsection{Scientific Foundation Models}
\label{sec:foundation-models}

Recent progress in AI has shown that large, flexible
parameterizations trained to simultaneously model large datasets across
different modalities (text, image, video, audio) can develop general, abstract
representations of the same object in different modalities.
This manifests as the ability to describe and reason about
image and video data, or to produce images and video from language inputs.
Applying this same functionality to physics data from experiments, observations, and
simulations could yield powerful ``scientific foundation models''~\cite{Golling:2024abg,Birk:2024knn,Hsu:2026sww,Mokhtar:2026gdb,DasBakshi:2025dcg,Parker:2025kne} whose
impact extends well beyond the custom, task-specific methods described above.

Beyond their direct utility, scientific foundation models are 
themselves useful objects of study, 
since they encode, in some fashion, a theory describing the physics data.
Interpretability and explainability techniques~\cite{Kriesten:2024are, Gambhir:2026cly} and
learned representations~\cite{Gambhir:2024dtf}, \textit{etc.},
may reveal characterizations of these implicit theories that are
comprehensible by humans.
It may be possible to 
derive useful effective theories, achieve more
detailed characterizations of experimental artifacts, or discover
new physics hidden in data that was missed by traditional analysis techniques.

More speculatively, physics data could be incorporated into
a generalized vision-language model~\cite{Sagar:2025mrd}.
Presently, both humans and AI
are constrained to clever data reductions to map
high-dimensional, heterogeneous physics data
into tractable summaries \cite{Franceschini:2022vck}.
However,
an AI model could natively ingest and reason about such data
without lossy intermediate reductions,
exposing powerful new capabilities and improving our ability to identify interesting
but subtle features in complex physics datasets.
From the AI perspective, particle physics
provides an unexploited reservoir of highly structured data to 
help improve AI's general reasoning capabilities.

\section{Building an AI Theory Community for the Genesis Mission}

We have argued that a large-scale collaboration among theorists, similar to the organization
in frontier experiments,
provides many shared assets: frontier-scale computing resources and access to frontier models, the large amounts of
data that constitute the agent-interaction corpus, shared validation infrastructure for benchmarking, and
scientific foundation models.
These shared assets are particularly valuable to theorists.   Individual theory groups typically do not command compute or model
budgets on the scale that experimental collaborations pool collectively (Section~\ref{sec:need-for-community}).
While theorists can share infrastructure with an experiment-oriented effort, there are
special considerations for AI-driven theory.  Theorists work differently with AI, often in long, exploratory, interactive sessions without a guaranteed 
outcome.   
This methodology does not fit well with batch-computing and allocations based on promised outcomes, a model that governs most existing computing infrastructure (Section~\ref{sec:hardware}).
We have further outlined the type of research theorists are doing and can do with AI.
While there is overlap with experimental research,
theory is often unique and valuable to particle physics as a whole.
To empower this research, theory needs to have a voice in the
organization of a field-wide AI effort, and that requires a community.

The national laboratory theory groups can serve as a backbone for building the larger collaborative units that we need to accomplish the type of vision presented here.
By employing staff scientists
and hosting and maintaining shared services such as the agent-interaction repository,
the national laboratories can 
provide the institutional continuity that individual university groups cannot easily sustain
against the mobility of graduate students and postdoctoral researchers.

We provide concrete steps for forming and developing this community in Section~\ref{sec:conclusions}.
Realizing this vision will require attention to access, infrastructure, policy,
and education.  
As the Genesis Mission Platform is made available, we expect that large, well-supported groups
centered at national laboratories will have the resources available to efficiently leverage it.
Lack of resources may be an entry barrier for individual researchers at universities, particularly
those who are less interested in AI infrastructure and more interested in research problems.
This pattern is consistent with past experience: during the development of the Leadership Computing Facilities (LCFs) at the national labs,
USQCD \cite{usqcd} played an important coordination role, enabling access for individual researchers to LCF resources and providing assurances, through the use of the USQCD Software API, that valuable LCF resources would be used efficiently. The proposed community organization will provide analogous services with the Genesis Mission Platform.

\subsection{Procurement and Model Access}
\label{sec:procurement}

There is no substitute for general, fluid intelligence in theoretical physics.
Theoretical physics proceeds by solving challenging problems, each solution
exposing new problems in turn.  To accelerate this process with AI, general
quantitative problem-solving ability is necessary, and there is accordingly no
substitute for access to the best models available at the frontier of AI
capabilities.  This has important implications for procurement.

Model capabilities are improving rapidly, and multiple major frontier
providers are competing to develop the best AI models.  A snapshot of the present
ranking of model capabilities is unlikely to extrapolate reliably beyond even
a few months.  
It is important that procurement decisions
are flexible enough to 
allow access to the best models at any given moment without locking into 
a particular model family or provider.

There are broad efforts to measure the overall intelligence and utility
of AI models by constructing benchmarks~\cite{Chung:2025nsd,Zhu:2025qnm,Faroughy:2026dkj,Miao:2026gmx,Kwa:2025metr,Islam:2026ona}.  These are important and
useful, but also highly limited and should not be over interpreted.  There are
well-known issues with the useful lifetime of any benchmark, which is limited
by both saturation as models improve and ``benchmaxing,'' or training to the
test, which furthermore complicates their interpretation.  More fundamentally,
a single scalar metric cannot usefully capture capabilities across the broad
skillset required for theoretical physics.  While it will be useful to further
develop theory-specific benchmarks, procurement decisions should not be made
solely on these grounds.  Instead, a broad range of models should be made
available to theorists, with an aggregation of their opinions as the primary
deciding factor on utility.
Benchmarks, nonetheless, inform these decisions in a different way, by roughly
establishing which class of models reliably automate a given task and therefore
where frontier access may not be required.

\subsection{Open-Weights Models}
\label{sec:open-weights}

Open-weights (ideally, open-source) models are obviously desirable from an
academic perspective.  From a procurement perspective, they can also be offered
at significantly reduced cost compared to frontier models. Open-weights models are
increasingly designed for efficiency.  Moreover, they can be hosted locally, on
either institutional hardware or the cloud, providing bare-minimum prices
modulo economies of scale.  Hardware purchased to serve these models can also
be used for other computational science tasks, providing additional benefits.
In principle, locally hosted models also allow more reliable guarantees of
information security: data is never streamed to any corporate entity.

However, at time of writing, open models are not an equivalent substitute for
frontier models and in practice are qualitatively less useful for general
theoretical physics tasks.  While open models developed by industrial and
academic efforts~\cite{grattafiori2024llama3herdmodels,openai2025gptoss120bgptoss20bmodel} are improving at similar rates to closed frontier models,
there is a well-known lag.  Common heuristics put it in the range of six months
to a year.  Further investment in open model development may close this gap.
However, if institutional decisions force theorists to rely solely on these
options, they will be disadvantaged.

\subsection{Modes of Access}
\label{sec:modes-of-access}

Virtually everyone
with an Internet connection has access to some
form of chatbot AI.  However, as already emphasized, access to the
best available models is critical for theory needs.  This is especially true in
the highly interactive chatbot mode, which delivers outputs directly to users;
low-quality outputs result in confusion and wasted human time.  At present
price levels, access to frontier models is affordable but nontrivial for
individual users.  For junior career researchers, it can be a substantial
financial burden.  Policies should be adjusted to fund this access.
Information security is also a concern.

Access to AI in the now-familiar ``chatbot'' form is important, but only
exposes a small fraction of the capabilities of emerging agentic use-cases.
Off-the-shelf proprietary agentic tools like Claude Code, Codex, and similar
products are proving to be extremely useful science tools~\cite{Schwartz:2026ekw,Badea:2026klb,Moreno:2026mqk}.
Enabling custom-built agentic AI requires API-level access~\cite{MCP,AISuite,Roman:2026rcn},
which can be substantially more expensive than subscription-based access.  More aggressive uses of
agentic AI cannot be expected to be paid for out-of-pocket by researchers and should be funded by grants and institutions.

\subsection{Hardware and Computing Infrastructure}
\label{sec:hardware}

Using agentic AI imposes hardware needs distinct from ``inference service''
hardware where the actual language model is run.  Modern agent systems are not
fully web-native in their most useful mode.  Instead, they read and write
files, interact with other data sources, and, critically, run programs and
execute code.  This requires a computing environment for the agents to work
within \cite{Menzo:2026xat}.

Much can be enabled simply with high-quality laptops or workstations.  For
junior-career researchers, this can be a significant financial burden.  At the
policy level, it will be important for grants to account for this in equipment
budgets.  At the group level, PIs should be encouraged to provide this
now-critical equipment to their graduate students and postdocs.

However, there are limitations to what can be accomplished on a laptop,
especially in computationally intensive pursuits.  There is a critical and
unfilled need to provide broad and easy access to larger-scale computing resources to
allow experimentation.  A combination of group and institutional clusters and
cloud computing provide the simplest solutions, but may be financially
prohibitive, and moreover have an upper bound on scale.

At present, allocations are granted at LCFs and
other supercomputing facilities based on promised outcomes.  This is
well-motivated for traditional computational science, but incompatible with
enabling AI/ML research.  AI/ML research generically requires computational
experiments with no guaranteed physics output: the experiment itself, and how
well it performed, is the useful output.  This includes both training of new
models, as well as automated computational science with agentic AI.  Most of
this type of R\&D is ``wasted compute'' in the framework of traditional
allocation structures, but is critical to make progress.  This is the role that
can and should be filled by the American Science Cloud (AmSC) \cite{amsc}, but should also
be addressed by updating allocation policies.
Efforts in this direction are already underway: the USQCD Collaboration, for
example, is currently engaged with the AmSC as an infrastructure partner and is
supporting seed AI model teams.

\subsection{Education and Workforce Development}
\label{sec:education}

Realizing the vision described in this document requires that theorists at all
career stages develop fluency with AI tools.  This represents a significant
shift in the skillset expected of theoretical physicists.
Not every theorist must become an AI/ML expert, but
a working familiarity with these tools (their capabilities and
limitations) should become standard.

At the graduate level, training should focus on
using AI tools for physics research and understanding the principles behind them.

For established researchers, the barrier is different: many are skeptical of or
unfamiliar with what AI tools can do.  This can be addressed with accessible
demonstrations, workshops, and community forums that showcase practical
applications relevant to their specific research areas, building on successful educational efforts primarily aimed at junior researchers~\cite{codas-hep, tac-hep, iaifi, iaifi-summer-school, ml4fp-2025, c2-the-p2}.

It is especially important to support junior-career researchers (graduate
students and postdocs) who may be the most natural early adopters but who also
face the greatest financial and institutional barriers~\cite{DOE-RFI-workforce-development, NSF25-349, ncses2021doctorate}.

Mentoring norms, on-boarding resources, career guidance and skill development, recognition and evaluation criteria, and other issues need to be discussed
in an open forum.

\section{Conclusions and outlook}
\label{sec:conclusions}

The use of AI in the natural sciences is a reality and an opportunity to
advance HEP theory while contributing to the general AI eco-system.
However, the exploitation of this opportunity requires structural change
in the theory community.   Theorists can still maintain the value of their
unique practice while finding ways to share resources.   The USQCD model has
demonstrated this can work.

Concretely, we propose the following near-term sequence. First, national laboratory theory groups convene an initial community meeting, before the end of 2026, to establish the scope of the organization based on inputs from both university and laboratory based groups: how frontier model access is pooled and procured, what shared benchmarks the community most needs, how agentic interaction data is formatted and governed, and what a community-scale foundation-model effort would require. Second, the participating labs create the first of these shared services, a pilot artifact repository on the Genesis Mission Platform to share agent-interactions, skills, and tools,
open to any contributing theory group regardless of institutional affiliation. We propose this service first, because it is the least expensive one to build, and its value grows fastest with the number of contributing groups. Third, these services are connected to the broader national effort through the Genesis Mission Platform, interoperating with the experimental community's corresponding infrastructure~\cite{Aarrestad:2026xrs} through common formats.
Such a structure will enable the HEP theory community to thrive in the AI era and tackle open physics problems that would be otherwise out of reach, and will aid in education and workforce development.
\section*{Acknowledgments}
The National Science Foundation and the Department of Energy provided support for individual contributors participating in this effort.
Support for this manuscript has been provided by the Fermi Forward Discovery Group, LLC under Contract No.~89243024CSC000002 with the U.S. Department of Energy, Office of Science, Office of High Energy Physics.
\printbibliography

\end{document}